\documentstyle[aas2pp4]{article}

\begin{document}
\title{Molecular Hydrogen and Paschen-alpha Emission in 
Cooling Flow Galaxies}
\author{Heino Falcke\altaffilmark{1}}
\affil{Astronomy Department, University of Maryland, College Park, MD
20742-2421 (hfalcke@astro.umd.edu)}
\altaffiltext{1}{Current Address: Max-Planck-Institut f\"ur
Radioastronomie, Auf dem H\"ugel 69, D-53121 Bonn, Germany}
\author{M. J. Rieke, G. H. Rieke}
\affil{Steward Observatory, University of Arizona, Tucson, AZ 85721
(mrieke,grieke@as.arizona.edu)}
\author{Chris Simpson}
\affil{Jet Propulsion Laboratory, MS 169--327, 4800 Oak Grove Drive,
Pasadena, CA 91109 (bart@fornax.jpl.nasa.gov)}
\author{Andrew S. Wilson\altaffilmark{2}}
\affil{Astronomy Department, University of Maryland, College Park,
MD 20742-2421 (wilson@astro.umd.edu)}
\altaffiltext{2}{Adjunct Astronomer, Space Telescope Science
Institute}

\begin{abstract}
We present near-infrared spectra obtained to search for Pa$\alpha$ and
molecular hydrogen lines in edge-darkened (FR\,I-type) radio galaxies
with bright H$\alpha$ emission in the redshift range $0.0535<z<0.15$.
We find that all three galaxies in our sample (PKS 0745-191, PKS
1346+26, \& PKS2322-12) which are associated with strong cooling flows
also have strong Pa$\alpha$ and H$_2$\,1--0\,S(1) through S(5)
emission, while other radio galaxies do not. Together with earlier
observations this confirms claims that cooling flow galaxies are
copious emitters of molecular hydrogen with large
H$_2$\,1--0\,S(3)/Pa$\alpha$ ratios in the range 0.5 to 2. The
emission is centrally concentrated within the inner few kiloparsec and
could come from warm (T$\simeq$1000-1500 K) molecular material which
is being deposited by the cooling flow. We speculate that the H$_2$
emission could be related to the interaction between the jets and this
molecular gas.
\end{abstract}

\keywords{galaxies: clusters: general -- cooling flows -- galaxies:
ISM -- galaxies: jets -- galaxies: ISM -- galaxies: elliptical and lenticular, cD}

\section{Introduction}
Cooling flows (hereafter CFs) are usually invoked to explain the huge,
extended X-ray luminosity seen in many massive elliptical galaxies and
clusters. They are thought to represent a massive, more or less
spherical, subsonic inflow of hot gas due to radiative cooling in the
central regions of the galaxy (e.g.\ Fabian 1994). The typical inflow
rates are of the order of several tens to several hundreds of solar
masses per year. However, it has been a puzzle where the huge amount
of inflowing matter is actually deposited after it has cooled, since
there was little direct evidence in any other waveband for this
material (e.g.\ CO or \ion{H}{1}, O'Dea et al.\ 1994, 1995).

Elston \& Maloney (1994) presented a brief report that strong
1--0\,S(1) $\lambda$2.1213$\mu$m molecular hydrogen emission occurs in
CF galaxies, however no individual detections are presented
in that paper.  Jaffe \& Bremer (1997, hereafter JB97) also have detected strong
H$_2$\,1--0\,S(1) emission in three CF galaxies (Abell 478,
Hydra-A, PKS0745-19) and none in a control sample of five galaxies,
suggesting that such emission may be characteristic of CF
galaxies. JB97 suggested that the molecular hydrogen
emission lines, which usually probe shocked or X-ray heated gas with a
temperature of 1000--2000\,K, trace some of this missing matter.

In this letter we report near-infrared spectroscopy of more CF
and control galaxies which for the first time also include transitions from
S(2) to S(5). The galaxies were observed in a program to search for hidden,
broad Pa$\alpha$ lines in FR\,I radio galaxies, to test the idea that the
optical and radio properties of FR\,I and FR\,II radio galaxies differ not
because of a different central engine, but because of a `closed' obscuring
torus in FR\,I's (Falcke, Gopal-Krishna, \& Biermann 1994). For this
program, we selected a number of bright radio galaxies with FR\,I or
amorphous radio structure and luminous H$\alpha$ emission in the redshift
range $0.0535<z<0.15$, so Pa$\alpha$ would be redshifted into the K-band
window. The galaxies were selected from the samples of Morganti, Ulrich, \&
Tadhunter (1992) and Owen et al.\ (1995).  Since CF galaxies tend
to be overluminous in emission lines compared to normal radio galaxies
(e.g.\ Baum 1992), our sample had a number of such objects. However, since
the sample was selected with another objective, it also includes suitable
control galaxies with similar properties to those with CFs.

\section{Observations}

Our spectra were obtained during two observing runs in April and
December, 1996, at the Multiple-Mirror-Telescope (MMT) (operated
jointly by the Smithsonian Astrophysical Observatory and the Steward
Observatory) with the FSPEC near-infrared long-slit spectrometer
(Williams et al.\ 1993). Most of the time the weather was not
photometric. The galaxies in our sample were observed in the K-band
with either a 75 l/mm (2nd order) or a 300 l/mm (1st order)
grating. The slit-width was $\sim$ 2.5 pixels
(0\farcs43\,pixel$^{-1}$), and guiding utilized an infrared camera
viewing the spectrometer slit. The low-resolution grating covered the
whole K-band range, while the wavelength range for the high-resolution
grating was chosen such that it covered the range between Pa$\alpha$
and H$_2$\,1--0\,S(2). PKS~0745$-$191 was observed at two different
central wavelengths to obtain a broader spectral coverage.


During the observation, sets of four exposures of the galaxies were
taken, each of which was offset by $\sim$7\arcsec\ along the
slit. Every 60--90 mins we observed a nearby calibration star in the
same manner. The spectra were sampled on the chip by 2--3 pixels per
resolution element along the spatial and the dispersion axes.  The
loci of the spatial and dispersion axes in each dataset were
determined by tracing the OH airglow lines and the continuum of the
calibration star in each of its four positions along the slit. The
exposures were then geometrically transformed so that the spatial and
dispersion axes became parallel with the axes of the array.

The data were first reduced following standard procedures for FSPEC (e.g.\
Engelbracht et al. 1996). We then re-reduced the data with a set of newly
written IRAF routines which allow an almost fully automated data
reduction and guarantee that all data sets, and specifically the OH airglow
subtraction, are handled in an identical manner. The general approach was
first to subtract the combined preceding and subsequent frames from the
raw-image, which effectively removes the dark current and most of the OH
airglow lines. Each background subtracted exposure was then flatfielded,
median smoothed with a box size comparable to the instrumental
resolution to correct for cosmic rays, and geometrically corrected. We
then centered the continuum in a given aperture and extracted a
spectrum of the object and of the residual OH airglow $\sim$14\arcsec\
offset from the galaxy in the same frame. Finally, the
residual-subtracted spectra were co-added with appropriate scales and
weights, divided by the calibration star (normalized to a black body
spectrum) and wavelength calibrated using the OH airglow spectrum.  We
found that the automated procedure yields spectra similar to
those from the standard procedures. Comparison of the two independent
reductions gives assurance that the spectra are free of severe artifacts.

\section{Results}

The resulting spectra of three of our seven galaxies are shown in
Figure~1. Due to the weather conditions during our observations and
the resulting lack of good photometric data, we did not attempt a flux
calibration and the flux densities are given in detector units. For
each spectrum we fitted the continuum with a polynomial and
added/subtracted the one sigma error per pixel derived from the
variations seen in the individual exposures during the co-adding
process (dotted lines in Fig.~1). This procedure gives a robust
estimate of the noise level in our spectra which, due to atmospheric
absorption troughs and strong OH airglow lines, can be a function of
wavelength.  The pixel-by-pixel error estimates allow one to
distinguish artifacts due to improper sky-subtraction from real
spectral features.  Parameters of the line identifications are given
in Table 1. Where line strengths are poorly measured, either due to
terrestrial atmospheric interference or because the line fell near the
end of the spectrum, the results are placed in parentheses in the
table.

In three cases (PKS~0745$-$191, PKS~1346+26, \& PKS~2322$-$12), all strong
CF galaxies, we have discovered strong Pa$\alpha$ and molecular
hydrogen lines. We did not find any evidence for either Pa$\alpha$ or H$_2$
in Abell~610, B2~0836+29, or B2~0915+32. Although these galaxies have
relatively luminous H$\alpha$ emission, they are not known to have strong
CFs.

The most extreme emitter of molecular hydrogen is PKS~2322$-$12, where the
H$_2$\,1--0\,S(3) line has twice the flux of Pa$\alpha$ and we also
detected H$_2$\,1--0\,S(4). In the other two CF galaxies,
PKS~0745$-$191 and PKS~1346+26, the H$_2$\,1--0\,S(3)/Pa$\alpha$ ratio is
roughly one half. PKS~0745$-$191 also shows emission from the
H$_2$\,1--0\,S(5) line at the end of the spectrum.
With the exception of this line, whose flux is very uncertain due to poor
determination of the continuum, all the line ratios we observe are
consistent with thermal excitation being the dominant process, with a
temperature $T=1000$--1500\,K. This temperature is similar to that found
by Mouri (1994) for NGC~1275, another CF galaxy.

It is known that CFs result in a detectable level of H$\alpha$
emission, thought to be excited by shocks or photoionization (Heckman
et al.\ 1989). However, since the differential extinction between
Pa$\alpha$ and the H$_2$ lines is negligible, this ratio is a more
reliable indicator of excitation and the relative strength of
molecular hydrogen emission than H$_2$/H$\alpha$. Nevertheless, here
we estimate the H$_2$~1-0~S(1)/H$\alpha$ ratios for our objects in
order to facilitate a comparison with JB97. Assuming
H$\alpha$/Pa$\alpha$ = 5 (Hill, Goodrich \& DePoy 1996) and a typical
ratio 1-0 S(3)/1-0 S(1) = 0.5, we determine H$_2$~1-0~S(1)/H$\alpha
\approx 0.2$ (up to 0.8 for the extreme case), in agreement with JB97.

From the width of the OH airglow lines, we determined that our final
spectral resolution (FWHM) in the 300 l/mm grating  spectra is 
$\sim30$\,\AA\ (410\,km\,s$^{-1}$ at $2.2\mu$m). The line widths we measure are
clearly broadened with an observed FWHM of 40--50\,\AA, which implies
an intrinsic line width of several hundred km\,s$^{-1}$ (after
subtraction of the instrumental resolution in quadrature).

For the 75 l/mm grating we achieved a resolution (FWHM) of
$\sim60$\,\AA\ (820\,km\,s$^{-1}$ at $2.2\mu$m) and the lines in
PKS~1346+26 are only barely resolved.  Only the width of Pa$\alpha$
seems to be large; however, this is probably an artifact due to the
combination of a strong OH line and a deep atmospheric absorption
trough around $2.0\mu$m as indicated by a sudden increase of
the noise in the red wing of the line. Also in PKS~0745$-$191 we find
a suspicious, extremely broad base to the Pa$\alpha$ line; however,
there is a very good chance that this is due to a conspiracy of noise
peaks.
 
The lines in PKS 2322$-$12 and PKS 0745$-$191 seem to be concentrated
in the inner 2--3\arcsec\ of the galaxy (FWHM), while the emission in
PKS~1346+26 seems to be more extended ($\sim4$--5\arcsec). Due to the
limited signal-to-noise ratio of the spectrum of the low surface
brightness emission in this galaxy and the small stepping increments,
which lead to a spatial cut-off, it is difficult to specify the extent
more precisely. We also do not see a variation of the
H$_2$\,1--0\,S(3)/Pa$\alpha$ ratio within the inner 3--4\arcsec\ of
our galaxies exceeding 15 per cent, which may indicate that the lines
are co-spatial.

\section{Summary \& Discussion}

Our detection of molecular hydrogen in two new CF galaxies and
of additional strong H$_2$ lines in a third, confirm the claims by
Elston \& Maloney (1994) and JB97 that CF galaxies are
copious emitters of near-infrared molecular hydrogen lines. Moreover,
this claim is further strengthened by additions to the list of control
galaxies without CFs where H$_2$ was not detected. 

There are now seven CF galaxies (NGC 1275, Abell~478, PKS~0745$-$191,
Hydra~A, PKS~1346+26, Cygnus A, \& PKS~2322$-$12) which share the same
properties: they have a) high estimated inflow rates of several
100\,M$_\odot$\,yr$^{-1}$, b) relatively strong radio sources, c)
luminous H$\alpha$ emission, and d) luminous H$_2$ emission (Table
2). The last property makes these galaxies outstanding --- no other
class of radio galaxies has been found to have such a strong molecular
hydrogen emission. In particular, of a total of eight (JB97 and this
paper) galaxies without CFs but similar in other respects, none
exhibit H$_2$ emission when observed to similar detection limits.  On
the other hand Table 2 demonstrates that for a complete subset of
Fabian's (1994) list of brightest cooling flow galaxies all CF
galaxies with radio and optical emission have been deteced in H$_2$.

In the detected galaxies the molecular gas is concentrated within the
central few kpc of the galaxy and on those scales is most likely
co-spatial with the H$\alpha$-emitting gas. Since the close link
between optical emission lines has already been established (Heckman
et al.~1989), indicating that perhaps shocks may play an important
role, and even correlated x-ray and optical filaments have been found
in CFs (Sarazin et al.~1992), it is conceivable that also the H$_2$
emission is directly related to the CFs. Molecular line ratios are
consistent with thermal excitation around $T\sim1500$K.

The theoretical interpretation for the connection between the cooling
flow and the molecular hydrogen is discussed in greater detail in
JB97. These authors find that the molecular masses cooling through
T$\simeq2000$ K in the CF are a factor $\sim$100 too low to account
for the H$_2$ emission and suggest that continuous reheating of the
gas by shocks or photoionization must take place. In this case only a
small fraction of the total gas deposited by the CF is enough to
account for the emission. However, the high H$_2$-to-H$\alpha$ ratios
we and JB97 find in the CF galaxies are difficult to explain with
current models.  JB97 suggest that this requires H$_2$ excitation by
either slow shocks or suprathermal secondary electrons produced by
X-ray photoionization of cold clouds.

It is attractive to associate this excitation solely with the action
of the CFs.  However, all the galaxies detected in H$_2$ have powerful
radio jets that might be another ingredient underlying these
observations. For example, Seyfert galaxies too have narrow emission
lines with $v\simeq400$\,km\,s$^{-1}$ and sometimes strong molecular
hydrogen emission. As shown by {\it HST\/} observations (e.g.\ Capetti
et al.\ 1996; Falcke et al.~1998), their emission line regions are
often concentrated in distinct features (filaments and bow-shocks)
indicative of a strong interaction between their radio jets and the
interstellar medium.

Ruiz, Rieke, \& Shields (1997) claim that jet-powered photoionizing
shocks are the most likely source of excitation for the H$_2$ emission
in Seyfert galaxies.  Two Seyferts (in addition to the CF galaxy
NGC~1275) in a sample of 28 showed ratios typical of CF galaxies:
MCG~8-11-11 and Mrk~6. Ruiz et al.~explain these observations by
attributing the luminous H$_2$ emission to excitation by jet-powered
shocks as they interact with molecular material. The kpc-scale radio
jets in Seyferts are often confined to the disk of the host galaxy,
and so in some extreme cases may encounter a region rich in gas which
provides an ideal working surface for the jet to excite molecular
hydrogen. It is therefore tempting to speculate that ellipticals, with
their larger-scale jets and relative dearth of gas, require an
external process such as a CF to transport large amounts of molecular
material to a location where it can interact with the jet. This jet
interaction scenario has been discussed in more detail already for
PKS~1346$+$26 to explain the radio-optical alignment in this galaxy
(McNamara et al.~1996). 

As an alternative one could imagine that the enhanced H$_2$ emission
seen so far is a consequence of ongoing mergers in the cD galaxies
rather than of the CFs. Further NIR spectroscopy of cD galaxies with
and without CFs and high-resolution imaging with NICMOS will therefore
be crucial to identify the origin of the strong molecular hydrogen
emission in these galaxies.

\acknowledgements This research was supported in part by NASA under
grants NAGW-3268 and NAG8-1027, by NSF grant AST9529190, and by DFG
grant Fa 358/1-1\&2. We thank Chad Engelbracht and Kevin Luhman for
providing useful information for the reduction of FSPEC data. We are
grateful to Chris O'Dea for interesting discussions on CFs and to the
2nd referee for a quick response.

\onecolumn

\clearpage
\begin{figure}
\plotone{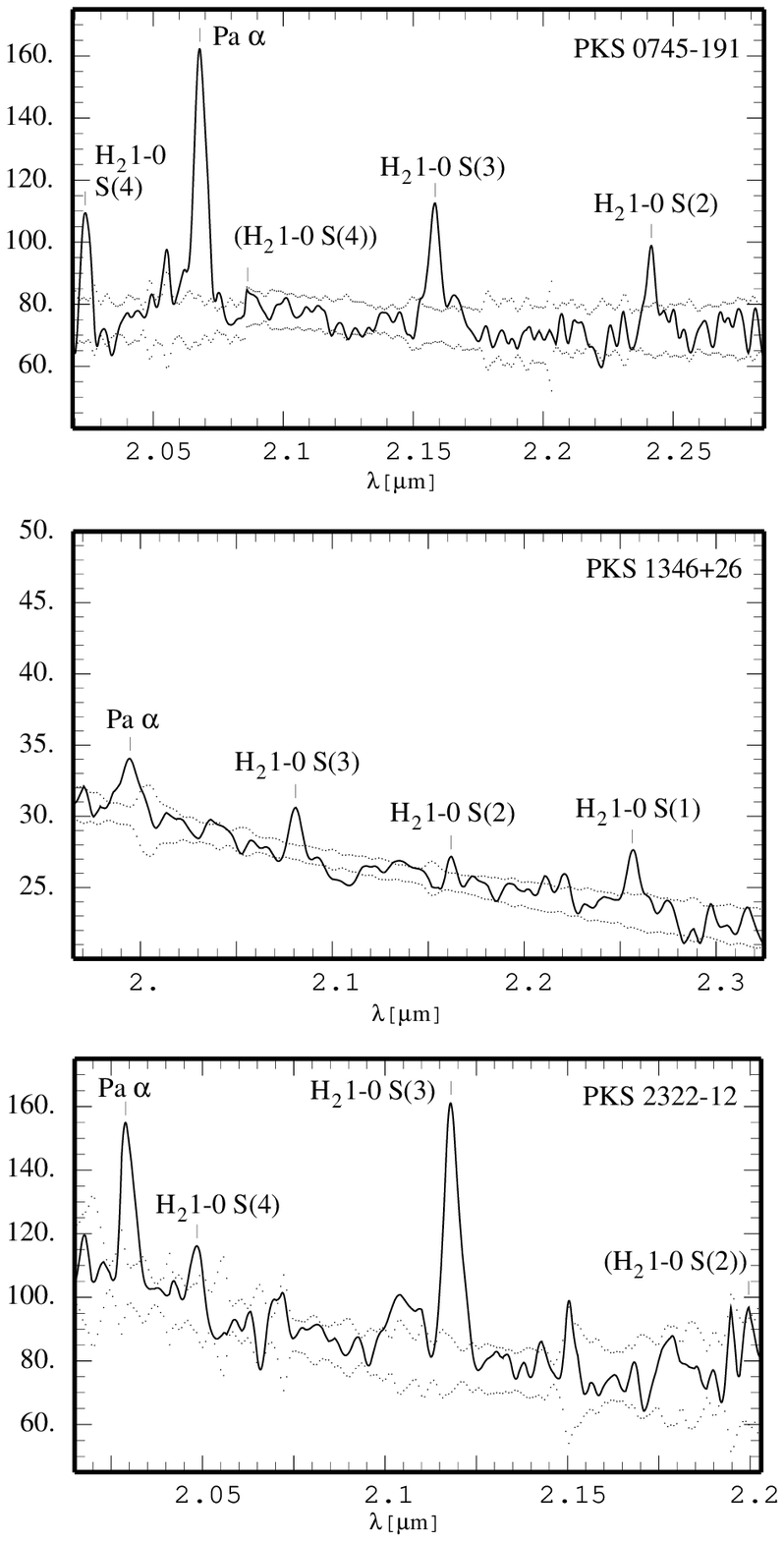}
\figcaption{\label{spectra} NIR spectra of galaxies with detected
H$_2$ lines in our sample plotted as $F_{\lambda}$ vs.~$\lambda$. The
flux values are in detector units. Dotted lines give the estimated
error derived from co-adding the individual spectra for each galaxy.}
\end{figure}








\clearpage
\renewcommand{\baselinestretch}{1}\small

\begin{deluxetable}{llllllll}
\tablecaption{Emission Line Parameters} 
\tablehead{ 
\colhead{Name}&
\colhead{$z$}&
\colhead{grating}&
\colhead{$\lambda$}& 
\colhead{flux}& 
\colhead{EW}& 
\colhead{FWHM}&
\colhead{Identification}\\
\colhead{}&
\colhead{}&
\colhead{[l/mm]}&
\colhead{[$\mu$m]}&
\colhead{}&
\colhead{[\AA]}&
\colhead{[\AA]}&
\colhead{}
}
\tablecolumns{8}
\startdata
PKS 0745$-$191& 0.1028 & 300 & 2.024 &  (0.46) & (31) & (42) & H$_2$ 1-0 S(5)
($\lambda1.8353\mu$m)\\
{}[CF]       &        &       & 2.068 &  1.00 & 55 & 52 & Pa$\alpha$\\
            &        &       & 2.086 & $\la$0.06&$\la$3.5& 53 &  H$_2$ 1-0 S(4)
($\lambda1.8914\mu$m)$^a$\\
            &        &       & 2.158 &  0.50 & 31 & 55 & H$_2$ 1-0 S(3)
($\lambda1.9570\mu$m)\\
            &        &       & 2.166 &  0.12 &  8 & 46 & Br$\gamma^b$ (or [SI VI])\\
            &        &       & 2.241 &  0.29 & 19 & 47 & H$_2$ 1-0 S(2)
($\lambda2.0332\mu$m)\\
\hline
Abell 610   & 0.0991 & 300  & 2.061 &  --   &$<$4&    & Pa$\alpha$\\
\hline
B2 0836+29  & 0.0643 &  75   & 1.996 &  --   &$<$7&    & Pa$\alpha$\\
\hline
B2 0915+32  & 0.062 &   75  & 1.991 &  --   &$<$5&    & Pa$\alpha$\\
\hline
PKS 1346+26 & 0.06326 &  75   & 1.995 &  1.00 & 17 &(129$^c)$ &  Pa$\alpha$\\
(Abell 1795)&         &       & 2.014 &  $\la$0.18&$\la$3& 67 &  H$_2$ 1-0 S(4)
($\lambda1.8914\mu$m)$^a$\\
{}[CF]        &        &       & 2.081 &  0.56 & 11 & 72 &  H$_2$ 1-0 S(3)
($\lambda1.9570\mu$m)\\
            &        &       & 2.162 &  (0.24) &  (5) & (52) &  H$_2$ 1-0 S(2)
($\lambda2.0332\mu$m)$^d$\\
            &        &       & 2.257 &  0.66 & 15 & 81 &  H$_2$ 1-0 S(1)
($\lambda2.1213\mu$m)\\
\hline
PKS 2322$-$12 & 0.0822 &  300  & 2.029 & 1.00 & 19 & 39 & Pa$\alpha$\\
(Abell 2597)&        &       & 2.048 & 0.32 &  7 & 32 & H$_2$ 1-0 S(4)
($\lambda1.8914\mu$m)\\
{}[CF]        &        &       & 2.118 & 2.02 & 51 & 50 & H$_2$ 1-0 S(3)
($\lambda1.9570\mu$m)\\
            &        &       & 2.200 & (0.63) & (18) & (46) & H$_2$ 1-0 S(2)
($\lambda2.0332\mu$m)$^a$
\tablecomments{
(1) -- Name ([CF] marks cooling flow galaxies), 
(2) -- redshift, 
(3) grating used in observation, 
(4) -- observed wavelength of line, 
(5) -- flux relative to Pa$\alpha$, 
(6) -- equivalent width in \AA{}ngstrom, 
(7) -- line width (FWHM) in \AA{}ngstrom (not corrected for
instrumental resolution), 
(8) -- line identification.
Notes: 
$^a$marginal detection,
$^b$insufficiently corrected Br$\gamma$ absorption in the 
calibration star, 
$^c$broad line width is most likely an artifact as indicated
by the sudden increase of the noise in the spectrum, 
$^d$after
subtraction of Br$\gamma$ in the calibration star.}
\enddata
\end{deluxetable}

\begin{deluxetable}{llrrrrl}
\tablecaption{Brightest Cooling Flow Galaxies with Radio
and H$\alpha$ Emission}
\tablehead{ 
\colhead{Name}&
\colhead{$z$}&
\colhead{$L_{\rm x}$}&
\colhead{$\dot M$}&
\colhead{lg P$_{408\rm MHz}$}&
\colhead{$L({\rm H}\alpha)$}& 
\colhead{H$_2$ 1-0}\\
\colhead{}&
\colhead{}&
\colhead{[$10^{43}$ erg/sec]}&
\colhead{[$M_\odot/{\rm yr}$]}&
\colhead{[erg/sec/Hz]}&
\colhead{[$10^{40}$ erg/sec]}&
\colhead{}
}
\tablecolumns{7}
\startdata
Abell 478	& 0.0861	&241	&570	&31.4&   19$^a$	&S(1)$^d$	\\
PKS 0745$-$191	& 0.1028 	&280	&702	&33.7&  490$^b$	&S(1)$^d$,S(2)-S(5)$^e$\\
Hydra A		& 0.0538	& 35	&315	&34.2&   34$^c$	&S(1)$^d$	\\
PKS~1346$+$26	& 0.06326 	& 89	&478	&32.6&  184$^c$	&S(1)-S(4)$^e$\\ 	
Cygnus A	& 0.05654	& 69	&187	&35.7& 2612$^c$	&S(1)\&S(3)$^f$\\
PKS~2322$-$127 	& 0.0822 	& 64	&480	&33.3&   65$^b$	&S(2)-S(4)$^e$
\tablecomments{Table of properties for galaxies from Fabian's (1994)
list of brightest cooling flow galaxies, for which we have applied our
original selection criterion ($z>0.0535$, as well as radio and optical
line emission) in hindsight.  With the exception of NGC 1275 (redshift
too low), this comprises the complete list of cooling flows with H$_2$
detections published so far. The columns description is: (1) -- name, (2) --
redshift, (3) -- 2-10 keV luminosity, (4) -- mass inflow rate, (5) --
log of monochromatic radio power at 408 MHz (from NED database), (6)
-- H$\alpha$ luminosity, (7) -- H$_2$ (1-0) lines detected so far.
Additional notes: 
Columns 3 and 4 are taken from Fabian (1994), columns 6 and 7 were
taken from:
$^a$ White et al. (1994),
$^b$ Zirbel \& Baum (1995),
$^c$ Godon et al. (1994),
$^d$ JB97,
$^e$ this paper,
$^f$ Ward et al. (1991). We use $H_0=50$ km sec$^{-1}$ Mpc$^{-1}$.
}
\enddata
\end{deluxetable}

\end{document}